\documentclass[conference]{IEEEtran}
\IEEEoverridecommandlockouts
\usepackage{cite}
\usepackage{amsmath,amssymb,amsfonts}
\usepackage{algorithmic}
\usepackage{graphicx}
\usepackage{textcomp}
\usepackage{xcolor}
\usepackage{multirow}
\usepackage{tabularx}
\newcolumntype{g}{X}                                \newcolumntype{s}{>{\hsize=.5\hsize}X}              \newcolumntype{Y}{>{\centering\arraybackslash}X}    
\usepackage{multicol}

\usepackage{hyperref}
\usepackage[aboveskip=4pt, belowskip=4pt]{subcaption}

\graphicspath{{figs/}{figures/}{pictures/}{images/}{./}} 

\usepackage{tabu}                      \usepackage{booktabs}                  \usepackage{lipsum}                    \usepackage{mwe}                       

\usepackage{mathptmx}                  \usepackage{listings}
\usepackage{tabularray}
\definecolor{codegreen}{rgb}{0,0.6,0}
\definecolor{codegray}{rgb}{0.5,0.5,0.5}
\definecolor{codepurple}{rgb}{0.58,0,0.82}
\definecolor{backcolour}{rgb}{0.95,0.95,0.92}
\definecolor{bluepoli}{cmyk}{0.4,0.1,0,0.4}
\lstdefinestyle{mystyle}{
	backgroundcolor=\color{backcolour},   
	commentstyle=\color{codegreen},
	keywordstyle=\color{magenta},
	numberstyle=\tiny\color{codegray},
	stringstyle=\color{codepurple},
	basicstyle=\ttfamily\footnotesize,
	breakatwhitespace=false,         
	breaklines=true,                 
	captionpos=b,                    
	keepspaces=true,                 
	numbers=left,                    
	numbersep=5pt,                  
	showspaces=false,                
	showstringspaces=false,
	showtabs=false,                  
	tabsize=2
}

\NewTblrEnviron{tableTesi}
\SetTblrInner[tableTesi]{ 
	width = \linewidth,
	colspec ={X[1]X[1]X[1]},
	columns ={l},
	hlines, vlines,
	row{1} = {bg=bluepoli, fg=white, font=\sffamily, c},
}

\begin{document}
	\title{A Gamified Framework to Assist Therapists with the ABA Therapy for Autism
		\thanks{Contact authors: 
			Matteo Cordioli (matteo.cordioli@mail.polimi.it),
			Dr. Laura Delfino (laura.delfino@solcomantova.it), Dr. Alessia Romani (aleromani24@gmail.com), and Dr. Elisa Mortini (elisa.mortini1@gmail.com).
			Pier Luca Lanzi (pierluca.lanzi@polimi.it; 0000-0002-1933-7717).}
	}
	\author{\IEEEauthorblockN{Matteo Cordioli$^{2}$, Laura Delfino$^{1}$, Alessia Romani$^{1}$, Elisa Mortini$^{1}$, Pier Luca Lanzi$^{2}$}\\
		\IEEEauthorblockA{$^{1}$Spazio Autismo Cooperativa Sinergie --- Mantova, Italy\\
		\url{https://www.solcomantova.it/servizi/autismo/}}\\
		\IEEEauthorblockA{$^{2}$Dipartimento di Elettronica, Informazione e Bioingegneria (DEIB) --- Politecnico di Milano, Milan, Italy}
	}

\maketitle
\begin{abstract}
We present a framework to assist therapists and children with autism spectrum disorder in their Applied Behavioral Analysis (ABA) therapy. The framework was designed in collaboration with Spazio Autismo, an autism center in Mantova, Italy. The framework is a first step toward transitioning from the current paper-based to fully digital-supported therapy. We evaluated the framework over four months with 18 children diagnosed with classic autism, ranging from 4 to 7 years old. The framework integrates a mobile app that children and therapists use during the sessions with a backend for managing therapy workflow and monitoring progress. Our preliminary results show that the framework can improve the efficacy of the therapy sessions, reducing non-therapeutic time, increasing patient focus, and quickening the completion of the assigned objectives. It can also support therapists in preparing learning materials, data acquisition, and  reporting. Finally, the framework demonstrated improved privacy and security of patients' data while maintaining reliability. \end{abstract}


\section{Introduction}
Autism is a neurodevelopmental disorder characterized by challenges in social interaction, communication difficulties, and repetitive patterns of behavior or interests. It is usually diagnosed in early childhood, and its exact cause is still not fully understood. A cure for autism does not currently exist; however, behavioral, developmental, educational, and psychological intervention strategies can significantly improve the lives of individuals \cite{hume2009increasing,CenterDiseaseControlTreatment}. 

In recent years, serious games have shown to be valuable potential as tools for aiding the treatment of individuals with autism \cite{HASSAN2021100417,doi:10.1080/10447318.2023.2194051}. This paper presents a gamified framework we developed to assist therapists and children with autism spectrum disorder during therapy. It has been designed in collaboration with Spazio Autismo in Mantova, a center specialized in supporting individuals of all ages with ASD. Its mission is to enhance the quality of life for individuals with autism and their families by promoting a cohesive and integrated approach. This is achieved through collaborative efforts, with a strong focus on working with families as the initial step towards creating a meaningful life plan. 

Spazio Autismo is staffed with a team of psychologists, consultants, and educators who act as case managers for the families of individuals with autism. The center applies the 
Applied Behavior Analysis (ABA) therapy \cite{skinnerIntroABA,ABA}, one of the most effective behavioral therapies that utilizes principles of learning theory to generate positive behavior changes and improve the quality of life for individuals with autism. 

Our framework has been conceived as a first step toward transitioning from the paper-based protocol currently followed at the center to a fully digital-supported therapy. It has been designed to align faithfully with ABA guidelines, support the entire flow of therapeutic sessions, help therapists create personalized sessions, and prepare its learning material. The framework comprises (i) an app for tablets for children to use during the sessions; (ii) a backend that records performance and behavior information; and (iii) a web frontend for therapists to plan sessions and review patients' progress.

We evaluated our framework over four months with 18 children diagnosed with classic autism, ranging from 4 to 7 years old. Our preliminary results show that time spent on non-therapeutic activities has been reduced with an increase between 10\% to 68\% of patients' focus. They also show a significant increase in the number of learning objectives completed compared to the paper-based protocol over the same period. In addition, the framework helped therapists collect more information about the sessions and patients' progress, reduce the time needed to prepare the learning material, and share the results with the families by automatically generating progress reports.  
\section{Autism and Applied Behavioral Analysis (ABA)}
\label{sec:aba}
Autism encompasses a spectrum of disorders, each with its unique characteristics \cite{lord2000autism}, and it manifests differently in each individual \cite{rutter1978diagnosis}. There are three main types of autism: Autistic Disorder, Asperger's syndrome, and pervasive developmental disorder not otherwise specified (PDD-NOS). Autistic Disorder, or classic autism, is characterized by significant impairments in communication, social interaction, and the presence of repetitive behaviors. Individuals with classic autism may exhibit delayed language development, difficulty forming relationships, repetitive movements, or fixated interests. This type of autism is often associated with intellectual disability. Asperger's Syndrome is considered a milder form of autism. Individuals with Asperger's may have average to above-average intelligence but struggle with social interaction and communication. PDD-NOS is a diagnosis often used for individuals with milder symptoms that do not meet the criteria for either classic autism or Asperger's syndrome.

\subsection{Applied Behavior Analysis (ABA)}
Applied Behavior Analysis (ABA) \cite{skinnerIntroABA,ABA} is one of the most effective behavioral therapies that utilizes principles of learning theory to generate positive behavior changes and improve the quality of life for individuals with autism and other developmental disorders. It is a data-driven, evidence-based approach that aims to bring meaningful and lasting behavior changes in individuals. By applying scientifically validated strategies, ABA helps individuals with autism and other developmental disorders to improve their skills, enhance their independence, and lead fulfilling lives \cite{efficacyInRecentYears}.

ABA involves systematic assessment and intervention strategies to target socially significant behaviors. It begins with a comprehensive evaluation to identify the individual's strengths, challenges, and specific behaviors that must be addressed. Based on the initial assessment, ABA interventions are tailored to each person's unique needs and abilities. Data collection and analysis are integral components of ABA, as they provide objective information about behavior patterns, progress, and the effectiveness of interventions.

\subsection{ABA Therapy Design}
The frequency of ABA sessions can vary depending on several factors, including the individual's needs, goals, available resources, and unique circumstances of each individual \cite{timeNeededABA}. ABA is considered a long-term intervention rather than a short-term fix. Consistency and continuity of therapy are important for skill development and behavior change. For some individuals, ABA may be needed for several years to address a range of skills and maintain progress over time. Integration of ABA principles into daily routines and activities is crucial to promote generalization of skills beyond the therapy sessions. ABA can be implemented across various environments, including homes, schools, clinics, and community settings, to support individuals in developing essential skills, such as communication, social interaction, self-help, and academic abilities.

\subsection{Verbal Behavior Milestones Assessment and Placement Program (VB-MAPP)}
\label{ssec:vb-mapp}
    The Verbal Behavior Milestones Assessment and Placement Program (VB-MAPP) is a criterion-referenced assessment based on typical language development, thus measuring how well an individual performs against an objective rather than another student. In practice, the tool provides an extensive checklist for tracking progress over time. 
    The VB-MAPP guidelines \cite{Montallana2019} suggest 18 categories, each with 15 objectives that must be completed in a predetermined order. 
    These categories are used to monitor the progress of ABA activities.
    
     In this study, we focused on three categories that the therapists identified as high priority: Tact, Listener, and Visual-Perceptual Matching to Sample (VP-MTS). \textit{Tact} is the ability to label or describe objects, actions, or events in the environment. It involves expressing what one observes using language.
	\textit{Listener} skills refer to receptive language abilities where an individual responds appropriately to verbal stimuli or instructions from others. It includes following directions and understanding spoken information. VP-MTS tasks involve matching or identifying objects, pictures, or symbols based on visual cues, often used for learning and discrimination tasks.
	 
\section{Related Work}
\label{sec:related}
There is a vast literature on the applications of serious game for autism spectrum disorders that has been recently surveyed in \cite{HASSAN2021100417,doi:10.1080/10447318.2023.2194051}. In this section, we overview the most relevant mobile applications available on the marketplace that are related to ABA therapy.

AutiSpark \cite{AutiSpark} is an educational mobile app designed for children with autism spectrum disorder that provides many engaging and interactive learning games tailored to meet the child's learning needs. It covers many topics, such as spelling, math, memory, sorting, matching, and puzzles. Its content spans various concepts such as image association, emotional comprehension, sound recognition, and much more. AutiSpark is exceptionally engaging and features a variety of vibrant colors and animated elements on the screen. This characteristic led the therapists to suggest that the app was designed for children who have already made some progress in their ability to discern between essential and non-essential elements.
Furthermore, the app lacks a clear differentiation between the various focus areas within ABA therapy. AutiSpark is designed as an app to entertain autistic children while ensuring they learn and enhance specific skills. However, it is not suitable to adhere to the ABA therapy guidelines.

Autism BASICS \cite{AutismBASICS} is a mobile app developed by a team of psychologists, speech, behavioral, and occupational therapists. It aims to engage children with autism and other special needs, enabling parents to work closely with their children.
Autism BASICS provides daily activities assigned by the child's therapist or parent, a library of learning-focused activities crucial for the development of children with autism, and a parent's corner with educational videos and content to facilitate collaboration between parents and their children. Parents or therapists assign tasks from the library, specifying the time the child should dedicate to these daily activities. Therapists can define multiple levels within a particular category for the child to focus on. Each day, the child completes these tasks presented in an engaging and enjoyable game format.
Autism BASICS offers thousands of activities categorized by significance in early childhood development for children with autism or developmental delays. These activities cover alphabet, spelling, math, and more, focusing on behavioral, academic, sensory, independent/self-help, communication, and social skills. According to the therapists at Spazio Autismo, some of these activities may be integrated into ABA therapy.

Autism ABC \cite{AutismABC} encompasses a range of essential features tailored to enhance learning and development in autistic children. It offers a selection of learning activities that cover cognitive, linguistic, and motor skills. The app is highly adaptable, making it suitable for children of various ages and autism spectrum levels. It ensures inclusivity by providing content that aligns with different developmental stages. The ABC App places a strong emphasis on data collection and analysis. It meticulously tracks and records each child's results for every activity, presenting them through user-friendly charts and indexes. These data-driven insights are valuable for assessing a child's progress. The app incorporates the concept of positive reinforcement, making learning engaging and enjoyable for children. This approach motivates children to participate in activities actively. However, therapists noted that it did not strictly adhere to the ABA therapy guidelines.

The applications we evaluated with the therapists have unique strengths but also areas for improvement with respect to Spazio Autismo's needs. AutiSpark impresses with its personalized learning approach and diverse content, yet it may benefit from increased data sharing and classification of games under the naming of ABA therapy. Autism BASICS excels in providing comprehensive content and user-friendliness but could enhance its adaptability and engagement. Finally, ABC App distinguishes itself with data-driven insights, though diversifying content and strengthening data anonymity could further improve its appeal. None of these apps faithfully align with ABA therapy, a major requirement for therapeutic use at Spazio Autismo. Accordingly, we designed a framework that could meet such strict therapeutic needs.

\section{The Framework}
\label{sec:framework}
Our framework was designed in cooperation with the therapists working at Spazio Autismo to help them transition from the current paper-based to a digitally supported therapy. The framework had to align with ABA guidelines faithfully, fully support the flow of therapeutic sessions, help therapists design personalized sessions, and preparing the learning material. The paper-based protocol makes it very difficult to record data about during the sessions. Accordingly, the framework had to record all available session data, store them securely, proving access using a web-based front-end with, possibly, some analytic support.

\subsection{Therapy Sessions}
Patients at Spazio Autismo have been diagnosed with \textit{classic autism}. They all have high support needs, which means they face significant challenges in social communication and interactions. Many of them may be non-verbal or have minimal speech, and they encounter substantial difficulties in performing daily tasks, such as self-care and vocational activities. Sessions are one-to-one (patient-therapist) meetings that last for approximately one hour. These sessions involve ABA activities that are closely monitored using the VB-MAPP categories (Section~\ref{sec:aba}).

Therapists interact with the patients with a deck of cards. Each card features a white background with a cartoonish image of an everyday object. The therapist presents these cards to the patient, selecting them based on the exercise being conducted and, when possible, the child's interests (e.g., cards with images of vehicles for children who show curiosity towards cars).

Each VB-MAPP category consists of 15 objectives of increasing difficulty. An objective is completed when the specified number of correct responses, as defined by the therapists, is attained. These values are determined for each category and each objective and are not dependent on the specific patient. During the session, the therapist only records the correct responses by marking them with checkmarks on an unstructured tracking paper sheet. Incorrect responses, including incorrect answers and unanswered questions, are not recorded.

Once the session is completed, the therapist has a sheet on which the correct responses obtained from the patient are written down. In preparation for the next session, the therapist reviews the sheet to avoid proposing the same responses the patient completed in the previous sessions. This tracking process must be repeated until that objective is completed. This tracking process can span up to several months.

Upon completing the objective, the therapist shares the correct responses with the supervisor and the patient's family. To do this, the therapist transcribes all the correct answers during the period required to complete the level into a form. The file is then printed and handed to the family.

\subsection{Framework Design}
The design of our framework focused on faithfully replicating the paper-based protocol already in place to maintain adherence to ABA principles and facilitate the transition to a fully digitalized protocol. Leveraging the technology's possibilities, therapists envisioned replicating the game experience of the deck of cards in a newly developed gaming app, introducing a gamified component. Therapists asked for tools that could support them in preparing the digital card deck used in the digital-based sessions. They also required the possibility of recording children's responses (with other performance and timing data) during the sessions and generating the reports shared with the family automatically. From the perspective of privacy and security, therapists must have access only to their patients' data. The children's identities must be kept anonymous accordingly, no personal identification data is stored on the backend.

\subsection{The Architecture}
\begin{figure}
	\centering
	\includegraphics[width=\columnwidth]{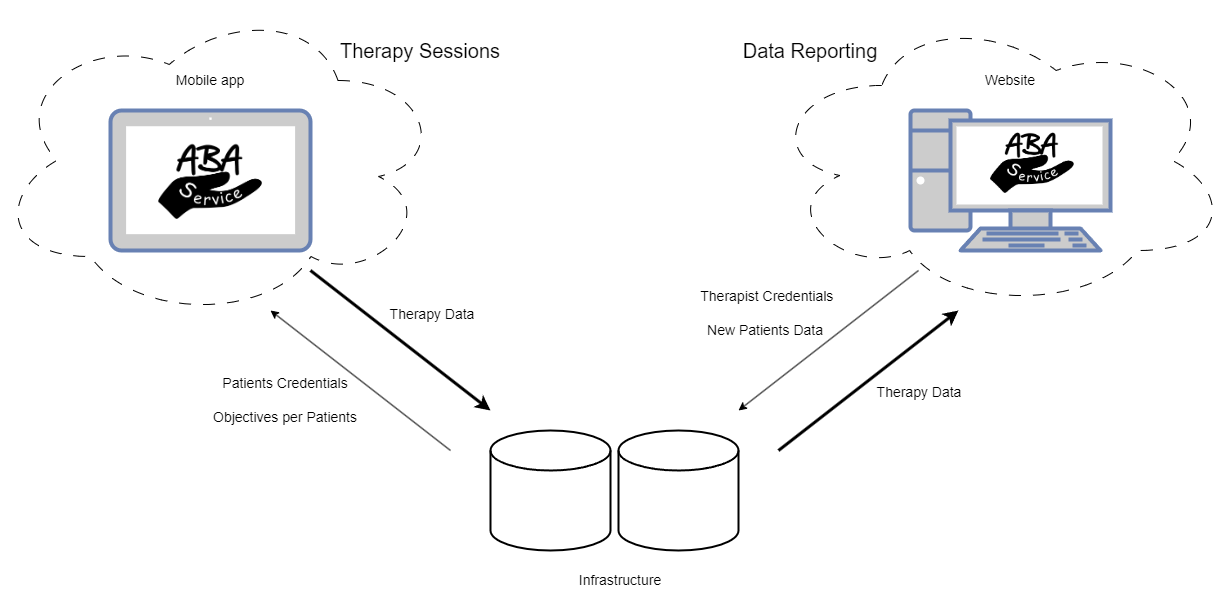}
	\caption{Data flow diagram}
	\label{fig:dataFlowDiagram}
\end{figure}

Figure \ref{fig:dataFlowDiagram} shows the architecture that comprises three components: the mobile app, a website, and the backend, enabling the communication between these two. The creation of the mobile app and the website became necessary to replicate ABA therapy in the digital format. The mobile app is exclusively intended for gaming by patients during their therapy sessions, and it is deliberately designed to be minimal and free of stimuli that could impact children's ability to maintain concentration in their therapy activities. Thus, the development of separate access via the web was necessary to provide the therapists with a space to manage data before meeting with the patients and after the session ended. 

At the beginning of each session, the therapist logs into the mobile app using the patient's credentials using a tablet. The therapist can access the patient data and view the planned and achieved objectives stored in the backend. While the patient plays ABA games on the tablet during a session, the app records the patient's performance data in the backend. Once the session is completed, the therapist can access the session data from the website. When the patient completes an objective, the therapist can download the report from the website to share with the supervisor and the family. The therapist can add or modify the patient's objectives from the website if necessary. At the moment of the evaluation, the framework has ABA games for three VB-MAPP categories (Section~\ref{ssec:vb-mapp}) that the therapists identified as high priority: Tact, Listener, and Visual-Perceptual Matching to Sample (VP-MTS). 

\section{Experimental Results}
\label{sec:results}
We evaluated our framework over four months with 18 children (7 females and 11 males) 4-7 years old following the ABA therapy at the Spazio Autismo in Mantova, Italy. Note that our evaluations included all the children who were eligible to participate; other children were excluded due to the specific requirements of their diagnoses. All the personal and therapeutic information about the children involved in the evaluation remained undisclosed. For research purposes, the therapists confirmed that (i) all the children involved in the evaluation could be collectively categorized under the diagnosis of classic autism; they all have high support needs, limited or no verbal communication abilities, and experience significant obstacles in their day-to-day activities. The patients are listed in the analysis with a unique numerical identifier. Our evaluation focused on three main aspects: \textit{Patient Focus During Therapy}, \textit{Speed of Objectives Completion}, and the \textit{Analysis of Error Rates}.

\subsection{Focus During Therapy}
\begin{figure}[t]
	\centering
	\includegraphics[width=.95\columnwidth]{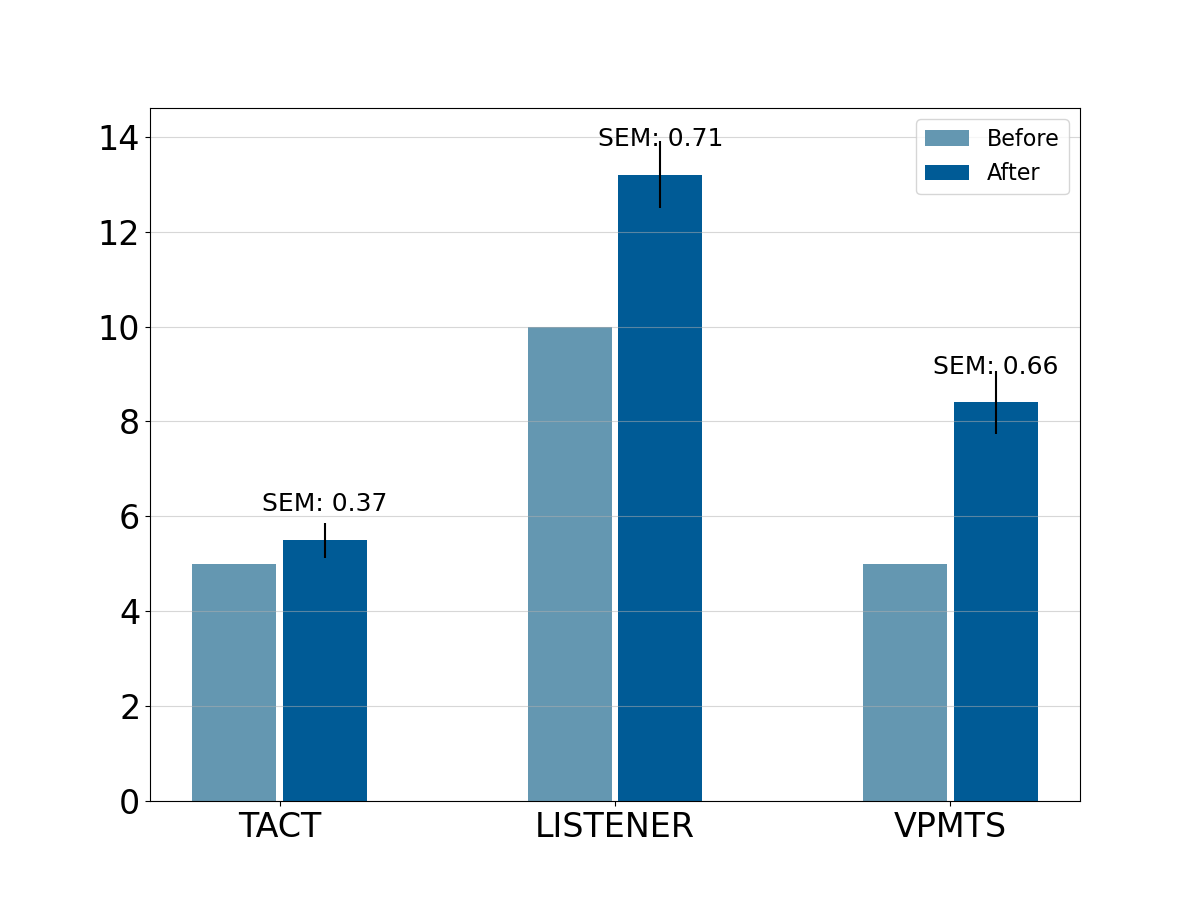}
	\caption{Average time spent per session in each category; bars report the standard error of the mean.}
	\label{fig:timePerSession}
\end{figure}

When using the paper-based protocol, children easily get distracted due to several stimuli in their learning environment. Our application cannot remove distractions entirely, 
however, therapists told us that children appeared more focused and engaged. We can estimate attention and involvement during sessions by quantifying the amount of time children spend actively engaged, computed as the difference between the last and first answer. Figure~\ref{fig:timePerSession} shows the average time previously spent using the paper-based protocol (column Before) and our framework (column After) in each category considered.
Note that, we do not report averages for the paper-based protocol since the therapists were not collecting timing for each session; the reported data are based on their own evaluation and session objectives.
For the \textit{Tact} category, our framework does not significantly increase the time spent actively engaged, and the role of digital support appears to be relatively neutral in increasing the focus during therapy. In the other categories, when utilizing the tablet as a supportive tool, children seem to become notably more engrossed and immersed in the therapeutic activities. We discussed the results with the therapists, who noted that \textit{Tact} is the category where the children do not actively operate on the tablet. Therefore, we recorded a slight difference between the paper-based and the digital supports. On the other hand, categories (\textit{Listener} and \textit{VP-MTS}) where children are more actively involved with the tablet during the sessions show a 32\% to 68\% improvement in the amount of time patients spend actively engaged. 

\subsection{Objectives Completion}
The time needed to complete therapeutic objectives (goals) set by the therapists is a critical indicator of improvement for ABA therapy. The children involved in this evaluation 
have different starting levels in different categories, as they may vary considerably depending on each patient diagnosis. Accordingly, our evaluation 
focused on the number of additional objectives the children could achieve in the given time frame. 

\begin{figure}[t]
	\centering
	\includegraphics[width=.95\columnwidth]{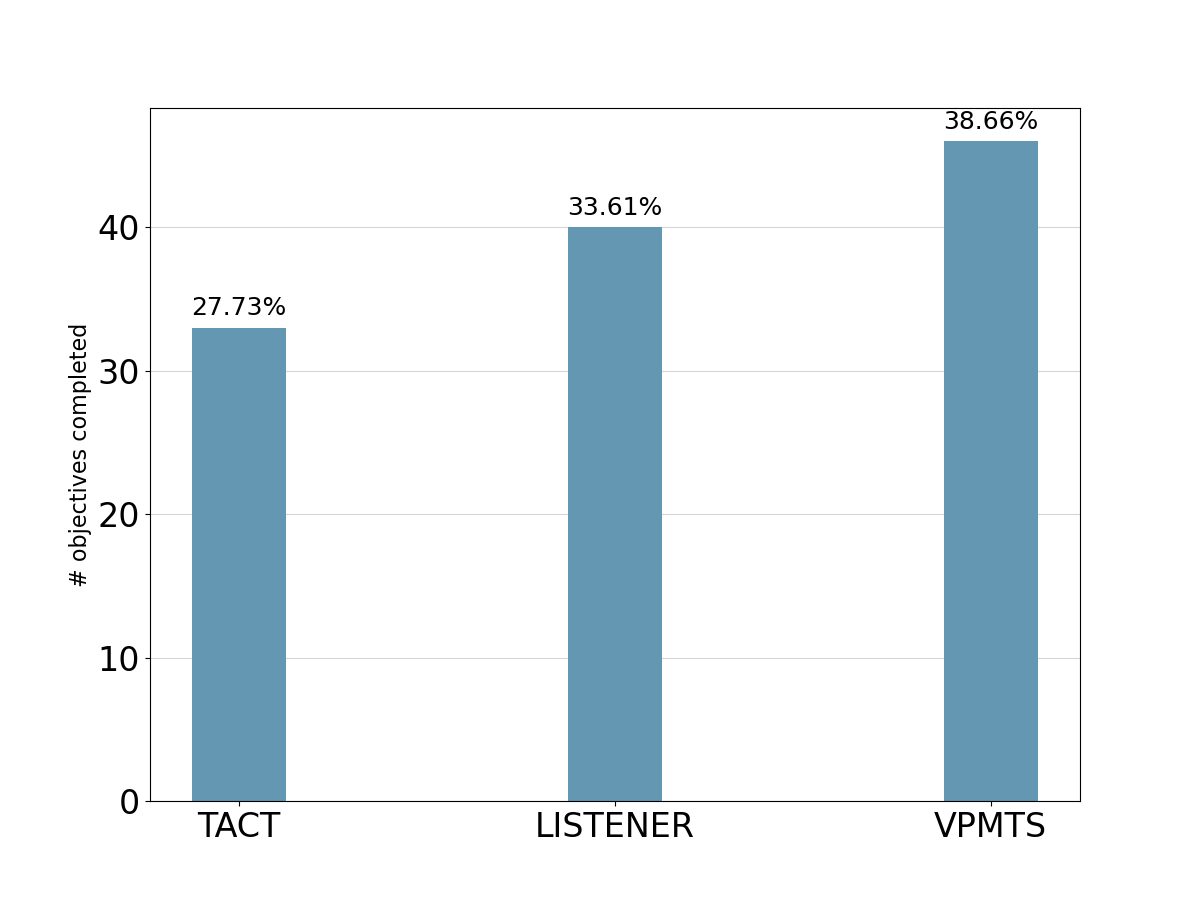}
	\caption{Percentage of completed objectives per category.}
	\label{fig:ObjPerPerCat}
\centering
	\includegraphics[width=.95\columnwidth]{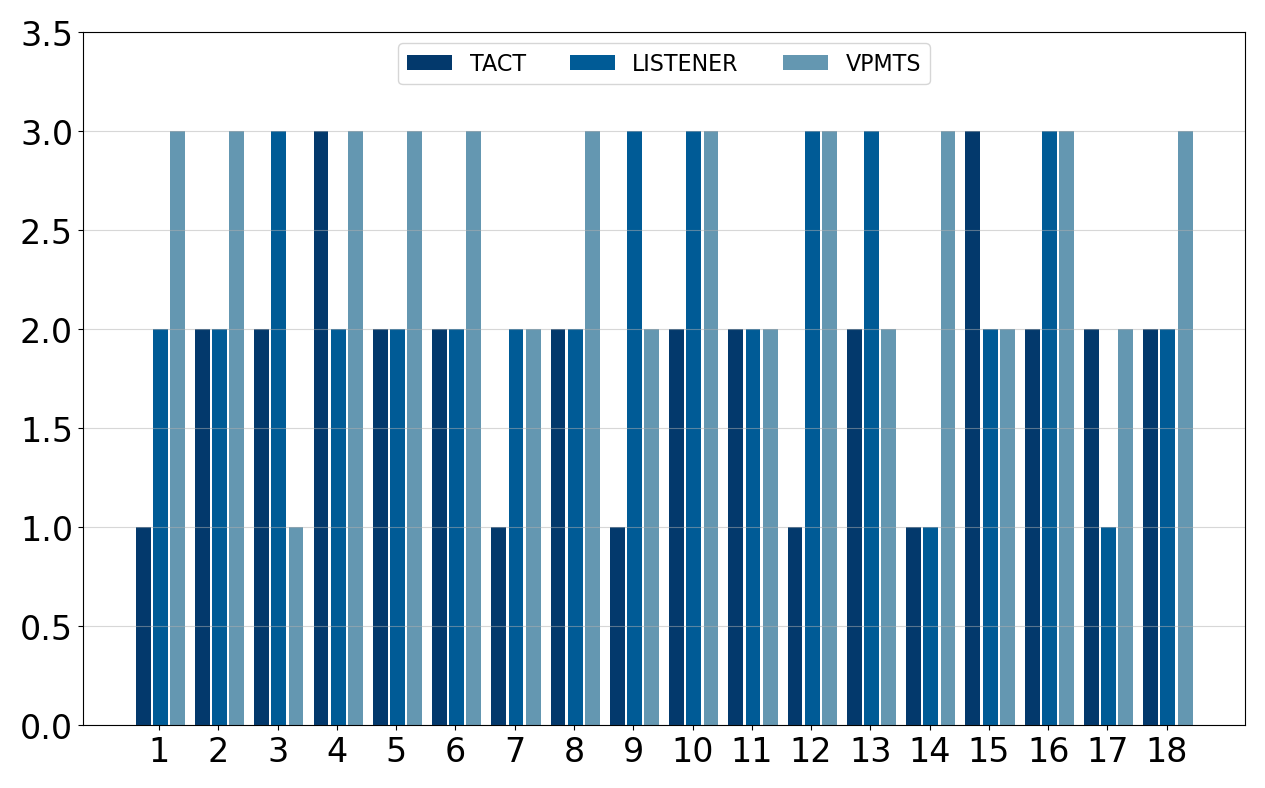}
	\caption{Number of objectives completed for each category by each patient.}
	\label{fig:NoObjPerCat}
\end{figure}

For each patient, we collected, the set of completed \textit{Tact},  \textit{Listener}, and \textit{VP-MTS} objectives. We also analyzed the (anonymized) paper reports collected for the same patients in the four months prior to the evaluation using the paper-based protocol; for consistency, we analyzed the same VB-MAPP. In the previous four months, 92 objectives were completed with an average of 5.11 objectives per children. At the end of the evaluation, 119 therapeutic objectives were accomplished using our framework with an average 6.11 objectives. Figure~\ref{fig:ObjPerPerCat} shows the average percentage of objectives completed for each category
while Figure~\ref{fig:NoObjPerCat} reports the number of objectives completed by each patient for each category. Note that the evaluation involved a limited number of patients at different phases of their therapy, of different categories, over a short time window. Accordingly, these results should be viewed solely as an assessment of a positive trend, also perceived by the therapists during the sessions.

\subsection{Analysis of Error Rates}
\begin{figure}[t]
	\centering
	\includegraphics[width=.95\columnwidth]{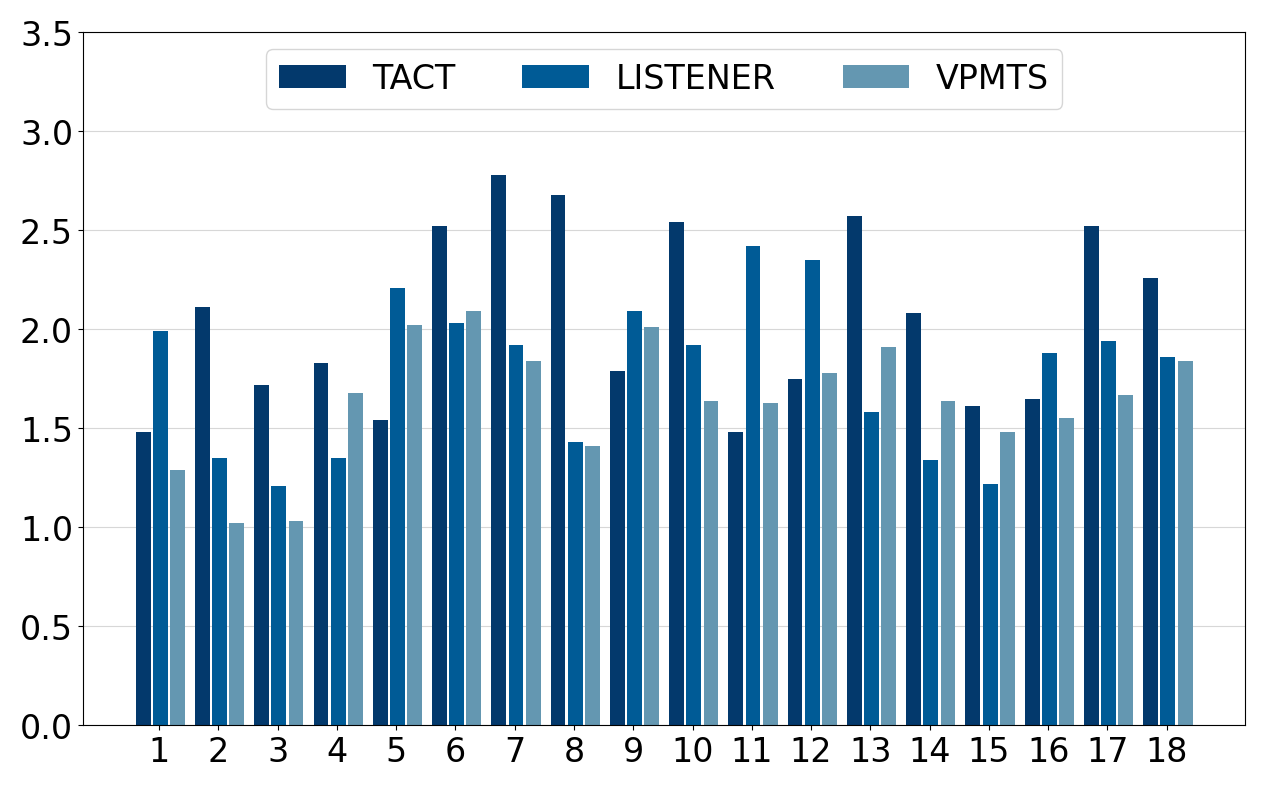}
	\caption{Error rate of each patient in each category.}
	\label{fig:errorDataGraph}
\centering
	\includegraphics[width=.95\columnwidth]{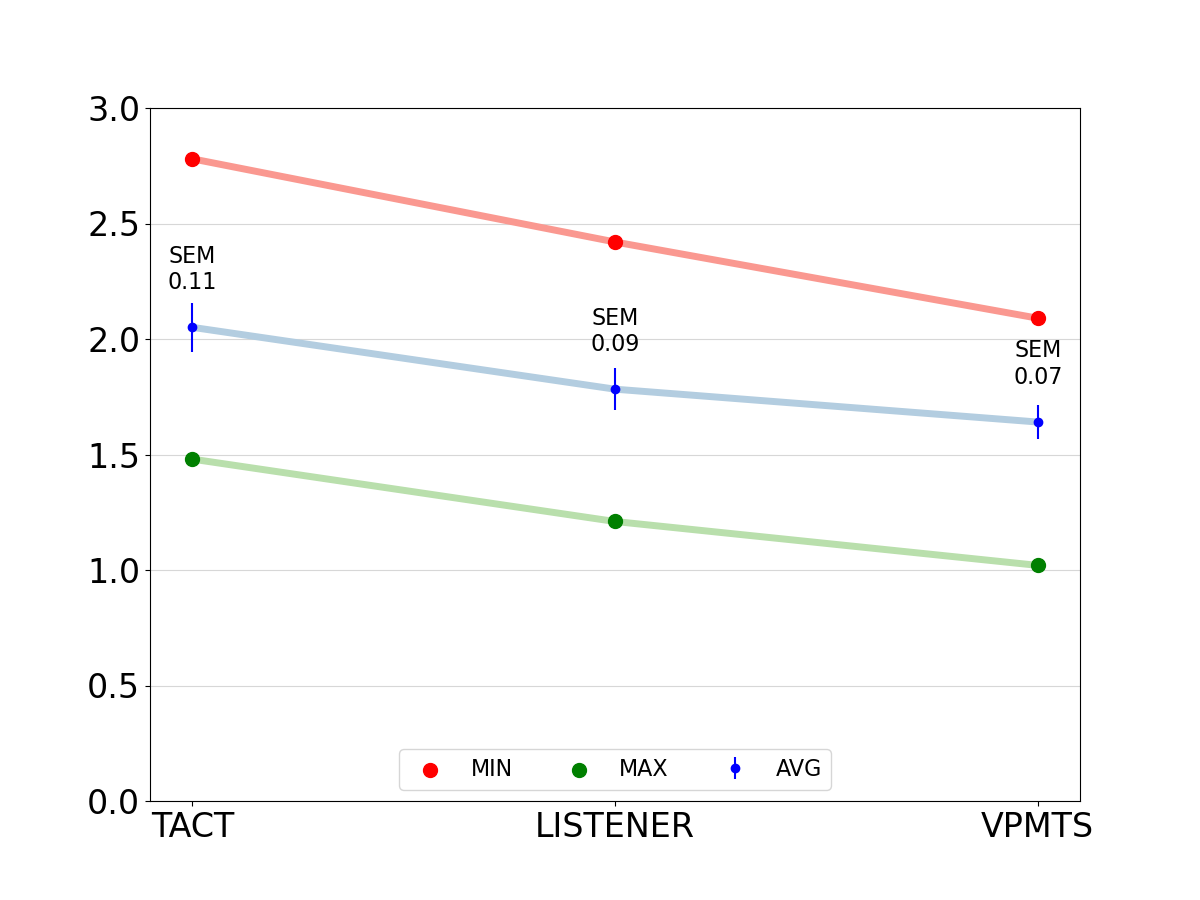}
	\caption{Error rate \(\psi\) per category: mean with standard error (blue line), minimum (green line), maximum (red line).}
	\label{fig:confErrors}
\end{figure}

Patients' errors were not recorded during paper-based sessions as they would distract the therapists from the patient. Our framework automatically records all the patient's interactions with the tablet, thus allowing therapists to analyze the number of errors and the error rate each patient makes during a session. We compute a patient's error rate as, 
\[
 	\psi = \frac{\# of Errors}{\# of Right Answers Required}
\]
If $\psi \leq 1$ then the patient made an equal or fewer number of errors compared to the correct answers; if $\psi > 1$, then the patient committed more errors than the required number of correct answers. It is worth noting that the objective is considered completed once the patient attains the prescribed quantity of correct answers, regardless of the number of errors made in the process. Figure \ref{fig:errorDataGraph} reports the average error rate for each category across all patients. 
The three categories show substantially different error rates. Figure \ref{fig:confErrors} shows the average error rate in each category with the corresponding range of variation. \textit{Tact} exhibits the highest average error rate (2.05), meaning that, for an objective requiring 50 correct answers, a patient would be presented with approximately $\approx150$ questions (50 answered correctly and 100 answered incorrectly). \textit{Listener} has an average error rate of \textit{1.78}, and \textit{VP-MTS} has \textit{1.64}. According to the therapists, the significantly higher average error for \textit{Tact} could be attributed to various factors. \textit{Tact} primarily focuses on enhancing the patient's ability to identify and name objects, actions, and events. This category often serves as one of the initial areas of intervention for patients who are new to ABA therapy and may encounter substantial challenges. Additionally, \textit{Tact} games typically lack the captivating elements or actions that can engage children, making it more challenging for them to maintain their focus during these activities. In conclusion, \textit{Tact} has exhibited the least favorable results from the perspective of patient outcomes among the three categories developed, coherently with the therapists' expectations, which had never been quantified before.

\subsection{Time Spent on Non-Therapeutic Activities}
Autism Spectrum Disorder (ASD) patients are prone to being distracted. Before the introduction of our app, many children lost their focus while therapists were setting up the learning and data collection materials. Our framework automates these two processes, and session setup is almost eliminated or reduced to the bare minimum. According to the therapists, our application has not only reduced their preparation and recording time but can also be considered an element of improvement in therapy efficacy.
 
\section{Conclusions and Future Developments}
\label{sec:conclusion}
We presented a gamified framework we developed to assist therapists in applying Applied Behavioral Analysis (ABA) therapy for autistic children. The framework includes an application for tablets that therapists use during therapeutic sessions with the children. The children's therapeutic activities are mini-games, providing safe and intuitive means to learn and improve in various categories. The application also offers therapists an efficient way to record several types of data automatically during the sessions. All the recorded data and the sessions' learning material can be accessed and prepared using a web app that provides access to the recorded data and some editing tools. We evaluated the framework at Spazio Autismo, with 18 children diagnosed with classic autism, ranging from 4 to 7 years old. Our preliminary results show that time spent on non-therapeutic activities has been reduced, the patients' focus increased by 10\% to 68\%, and speed of objectives completion accelerated, with an average of 6.61 objectives completed in 4 months, in contrast with the 5.11 before the digitalization. 

ABA therapy is heavily data-driven and the paper-based approach has posed significant challenges. Our framework helps therapists collect more data that can become integral to their analyses of patients' progress. For example, therapists could tailor the therapy to each patient, something infeasible with paper-based therapy. Parents could use our framework outside the sessions to offer children short, engaging games anytime during the day. This possibility fits patients undergoing ABA therapy, which relies on the repetition of behaviors.

\end{document}